\documentclass[prd,aps,preprint,amsmath,nofootinbib,amssymb,superscriptaddress,floatfix]{revtex4-1}
\usepackage{graphicx}
\usepackage{subfigure}
\usepackage{hyperref}

\begin{document}

\title{Holographic Entanglement Entropy and Complexity for D-Wave Superconductors}

\author{Yuanceng Xu}
\email{xyc@mails.ccnu.edu.cn}
\thanks{Corresponding author}
\affiliation{Institute of Astrophysics, Central China Normal University, Wuhan, Hubei 430079, China}

\author{Yu Shi}
\affiliation{School of Physics, Electronics and Intelligent Manufacturing, Huaihua University, Huaihua, Hunan 418000, China }

\author{Dong Wang}
\author{Qiyuan Pan}
\email{panqiyuan@hunnu.edu.cn}
\affiliation{Key Laboratory of Low Dimensional Quantum Structures and Quantum Control of Ministry of Education, 
Synergetic Innovation Center for Quantum Effects and Applications, and Department of Physics, Hunan Normal University, Changsha, Hunan 410081, China}

\date{\today}

\begin{abstract}
	Using the RT formula and the subregion CV conjecture, we numerically investigate the holographic entanglement entropy (HEE) and holographic subregion complexity (HSC) for two holographic d-wave superconducting models with backreactions. 
	We find that the HEE and HSC can probe these two d-wave superconducting phase transitions. 
	The HEE of the superconducting phase is always lower than that of the normal phase.
	For the HSC, however, it behaves differently and interestingly, which depends on both the strip-width $L_{x}$ and backreaction $\kappa $.
	More specifically, when the backreaction is larger than a particular critical value, 
	or the strip-width of the boundary subsystem is smaller than a particular critical value, 
	the HSC in the superconducting phase is larger than that in the normal phase, and vice versa.
\end{abstract}
\maketitle

\section{Introduction}

The anti-de Sitter/conformal field theories (AdS/CFT) duality (or gauge/gravity duality)
provide a powerful tool for strongly coupled quantum systems.
The gauge/gravity duality originated from the study of superstring theory.
More specifically, Maldacena proposed that the Type $IIB$ string theory ($D3$ branes) on $(AdS_5\times S^5)_N$ with some
appropriate boundary conditions (and possibly some boundary degrees of freedom) 
is dual to $4d$  $\mathcal{N}=4$  $U(N)$ Super-Yang-Mills theory in the large $N$ limit 
\cite{Maldacena,Gubser:1998bc,Witten:1998qj}.
In the past two decades, gauge/gravity duality has been widely applied to various fields of physics.
Two more well-known and successful areas are quantum chromodynamics (QCD) \cite{Aharony:1998xz,Karch:2000gx,Karch:2002sh} 
and condensed matter physics (CMP) \cite{Hartnoll:2009sz,Herzog:2009xv,McGreevy:2009xe}.
This paper mainly introduces the holographic duality of superconductivity in condensed matter physics. 
Since the first discovery of high-temperature superconductors (HTSC) \cite{Bednorz:1986tc} in 1986, 
people have tried to propose various theories for their electron pairing mechanism. 
Even the famous BCS theory has failed to provide a reasonable explanation for the HTSC systems with strong coupling.
As a new powerful tool, the gauge/gravity duality \cite{Maldacena,Gubser:1998bc,Witten:1998qj} 
can be used to study strongly coupled systems.
In 2008, Gubser pointed out the spontaneous $U(1)$ symmetry breaking by bulk black holes, 
which can be used to investigate the superconductor/conductor phase transitions in the dual CFTs \cite{Gubser:2008px}.
Then Hartnoll,Herzog and Horowitz built the first holographic superconducting model \cite{Hartnoll:2008vx}. 
Indeed, using the gauge/gravity duality, 
such a simple model can yield condensed curves similar to the results of BCS theory.
In addition to the holographic duality of the s-wave superconductor mentioned above, 
the authors \cite{Gubser:2008zu,Gubser:2008wv} established
a holographic duality of the p-wave superconductor by putting a $SU(2)$ Yang-Mills field around the AdS black hole.
However, most previous studies focused on the s-wave and p-wave order parameters. 
Many experiments \cite{Kotliar,Bickers,Moriya,Monthoux} have found that most HTSC materials are d-wave superconductors, 
meaning that the orbital angular momentum of the electron pair (Cooper pair) equals two.
Because of a possible relationship with cuprates, it would be an essential but challenging task to construct a complete 
and self-consistent (no other unphysical degrees of freedom) holographic dual theory of d-wave superconductivity.
 Chen,Kao,Maitly,Wen and Yeh (CKMWY) proposed a truncated model with sufficient ingredients to catch the main features of d-wave superconductors \cite{Chen:2010mk}. 
They replaced the scalar field in holographic s-wave superconductors with a tensor field. 
A d-wave state could be naturally described by a $3\times 3$ symmetric traceless tensor with five components. 
When the time component was considered, this symmetric traceless tensor could be expressed as
$B_{\mu \nu } (\mu ,\nu=0,1,2,3) $ with $ B_{\mu \nu }=B_{\nu \mu }$ and $B^{\mu }_{\mu }=0$.
Although this toy model has some features of d-wave superconductors, 
it's still an incomplete theory and contains spurious degrees of freedom that may cause instability. 
Subesquently, this model was further studied \cite{Zeng:2010vp,Zeng:2010zn,Li:2014wca,Guo:2014wca,Lin:2018ebe,Ge:2012vp}.
Inspired by the early studies of a neutral, massive spin-two field in a flat background \cite{Fierz},
Benini, Herzog, Rahman and Yarom (BHRY) \cite{Benini:2010pr} found a toy model for a charged 
spin-two field in an asymptotical AdS geometry which could be used to describe the d-wave order parameter.
Fortunately, the action they wrote down has the correct number of propagating degrees of freedom and is ghost-free and stable.
In the context of holographic superconductors,
it was often worked in the probe limit \cite{Hartnoll:2008vx} to simplify the calculations 
and catch the typical condensation properties.
However, sometimes we have to consider the effect of a full backreaction on the metric 
when calculating some particular quantities like the HEE and the HSC in holographic superconductors.
In fact, for the CKMWY d-wave model, the holographic superconductor with the backreactions has been studied, such as in \cite{Ge:2012vp}.
For the BHRY d-wave model, Kim and Taylor constructed top down models for holographic superconductors and suggested that the backreacted of the condensate
can be computed directly by working with higher dimensional theory \cite{Kim:2013oba}.
Thus, as an attempt, we consider the backreaction for the BHRY d-wave model in this work, without looking in detail at the constraint equations required
to obtain the correct number of propagating degrees of freedom, just as in the CKMWY d-wave model \cite{Chen:2010mk}.
For further research on this d-wave model the reader may refer to these articles \cite{Zeng:2010fx,Chen:2011ny,Gao:2011aa,Krikun:2013iha,Nishida:2014lta,Krikun:2015tga}.
One motivation for this paper is, when keeping both the conformal dimension $\Delta_{+}$ and backreaction $\kappa $ 
corresponding to two d-wave superconducting models the same, respectively, 
to compare the critical temperature of the phase transition between these models.

In addition to its wide application in CMP and QCD, 
the gauge/gravity duality has been recently extended in quantum information physics (QIP),
such as the entanglement entropy and the computational complexity.
Considering a quantum system divided into two subsystems, $A$ and $B$,
the total Hilbert space can be written as the direct product of two subspaces
$\mathcal{H}_{tot}=\mathcal{H}_{A}\otimes \mathcal{H}_{B}$.
Then the entanglement entropy of subsystem A can be defined by von Neumann entropy,
$S_{A}=-tr_{A} \left( \rho_{A}\log \rho_{A} \right)$ with the reduced density matrix $\rho_{A}=tr_{B} \left(\rho_{tot} \right)$.
The entanglement entropy can be used to measure how closely entangled (or how ``quantum'') a given wave function is.
It's very difficult to calculate the entanglement entropy of arbitrary submanifolds $A$ directly 
in higher dimensional QFT.
Fortunately, a holographic interpretation of entanglement entropy in CFT was proposed 
by Ryu and Takayanagi \cite{Ryu:2006bv,Ryu:2006ef}
\begin{eqnarray} 
S_{A}=\frac{Area(\gamma _{A})}{4G_{N}}    \label{HEE},
\end{eqnarray}
where $Area(\gamma_{A} )$ is the area of the static minimal surface in AdS spacetime whose boundary is given by $\partial A$.
$G_{N}$ is the Newton constant.
Indeed, the authors \cite{Ryu:2006bv,Ryu:2006ef} investigated the area of 
a one-dimensional minimal surface (i.e. the length of a geodesic) on the $AdS_{3}$ boundary,
and the entropy obtained by the holographic interpretation is precisely consistent with the result of $CFT_2$,
i.e. $S_{A}=\frac{c}{3}\cdot \log\left[\frac{L}{\pi a}\sin\left(\frac{\pi l}{L} \right) \right] $, 
even including the (universal) coefficients. Inspired by these two papers \cite{Albash:2012pd,Cai:2012nm}, 
where the HEE in s-wave and p-wave superconductors 
was studied by Albash and Cai et al., respectively,
in this paper, we would like to investigate the HEE in holographic d-wave superconducting models.
As mentioned above, most high-temperature superconducting materials are d-wave superconductors.
Therefore, it is necessary to investigate whether the
entanglement entropy can be used to probe d-wave superconducting phase transitions.
If so, then we can say that the HEE can be used to detect \emph{all} superconducting phase transitions.

Another information-theoretic quantity, 
computational complexity \cite{Susskind:2014rva}, was related to a gravitational concept 
within the context of AdS/CFT correspondence.
Computational complexity measures how difficult it is to carry out a unitary operation $U$.
More specifically, it is defined by the minimum number of ``basic'' unitary operations (``gates'') 
required to implement the operation $U$. 
For example, according to the AdS/CFT duality, the state $\vert \psi \left(t_{L},t_{R}\right)  \rangle  $ 
on the boundary of the analytic eternal two-sided AdS-Schwarzschild black hole
is determined by the thermofield double $\vert TFD \rangle  $ state.
The CV-conjecture (Complexity/Volume conjecture) \cite{Susskind:2014rva,Stanford:2014jda} proposes that
the complexity of the state  $\vert \psi \left(t_{L},t_{R}\right)  \rangle  $
can be obtained by the maximum spatial volume $V$ of ERB (Einstein-Rosen bridge)
connecting the boundary of the left and right black hole.
Another improved conjecture, the CA-conjecture (Complexity/Action conjecture) 
\cite{Brown:2015bva,Brown:2015lvg} proposes that
the complexity of a state is given by the classical action $\mathcal{A}$ of a region which is called 
the Wheeler-DeWitt(WDW) patch in the Penrose diagram.
Unlike the HEE, which exists only in the bulk geometry outside of the black hole, 
the holographic complexity (HC) allows us to delve into the inside of the black hole 
and reveal the physics behind the event horizon.
However, motivated by the HEE and the HC,
the volume enclosed by the minimal hyper-surface and infinite boundary in bulk may also define a complexity 
which is called holographic subregion complexity (HSC) \cite{Alishahiha:2015rta}
\begin{eqnarray}
	\mathcal{C_A}=\frac{V(\gamma_A )}{8\pi LG_{N} } \label{HSC},
\end{eqnarray}
where $L$ is the AdS radius, and $G_{N}$ is Newton's constant. 
It is worth noting that, due to the metric being divergent on the infinite boundary of an asymptotic AdS geometry,  
to regulate this quantity, we need to put a UV cutoff $\epsilon=\frac{1}{r}$ near the boundary $r=\infty$
just like we did in the HEE.
One can find the universal term of the HSC $C_{u}$ by subtracting the diverging term from the total complexity.
The HSC in s-wave and p-wave superconducting models has been studied
\cite{KordZangeneh:2017zyy,Chakraborty:2019vld,Fujita:2018xkl}.
In this paper, we want to investigate the HSC defined in Eq.()\ref{HSC}) above in d-wave superconductors.
 
Another research motivation of this paper comes from analyzing some existing results in the literature 
\cite{KordZangeneh:2017zyy,Yang:2019gce,Chakraborty:2019vld}. In Ref. \cite{KordZangeneh:2017zyy},
the authors claimed that the difference between the HEE (or the HSC) of the superconducting phase and that of the normal phase 
becomes more significant as the value of backreaction $\kappa$ increases.
Their investigation clarifies the conflict and shows that the HSC does not behave similarly to the HEE.
More specifically, they found that the HEE $S_{u}$ increases with $T/\mu $, 
while the HSC $C_{u}$ decreases with the increase of $T/\mu $.
Indeed, our results also show that the HEE does have a similar property, 
but the behavior of HSC is more complicated and determined by the subregion's strip-width $L_{x}$ and the backreaction's value $\kappa $.
Although the authors \cite{KordZangeneh:2017zyy} studied one-dimensional holographic s-wave superconductors,
we believe these two models are still comparable because the equations of motion of the d-wave models 
are similar to those of the s-wave.
Another thing worth noting is that in Ref. \cite{KordZangeneh:2017zyy},
the HSC of the superconducting phase is larger than that of the normal phase and decreases with the temperature increase. 
However, the authors of Ref. \cite{Yang:2019gce} claimed that the superconducting phase 
always has a smaller complexity than the normal phase below the critical temperature.
Although the latter considers the general CV-conjecture (Complexity/Volume conjecture) \cite{Susskind:2014rva,Stanford:2014jda}, 
it seems that something ambiguous about the behavior of complexity needs further investigation.
Our results show that when the backreaction $\kappa$ is larger than a particular critical value, 
or the strip-width $L_{x}$ of the boundary subsystem is smaller than a particular critical value, 
the HSC in the superconducting phase is larger than that in the normal phase, and vice versa, although
the physical mechanisms involved are not well understood.

This paper is organized as follows: 
In Sec. \ref{sec2},  
we briefly review the two holographic d-wave superconducting models respectively
and obtain the critical temperatures of phase transition under three different values of the backreaction.
In Sec. \ref{sec3}, we study the HEE and HSC of two holographic d-wave superconducting models.
In Sec. \ref{sec4}, we summarize and discuss our results.

\section{ D-wave superconductors} \label{sec2}
\subsection{CKMWY D-Wave Superconductor}

In this section, we will briefly introduce the d-wave superconductor described by
the background geometry of a black hole 
coupled with a symmetric, traceless second-rank tensor field $B_{\mu \nu }$ 
and a U(1) gauge field.  
The total action takes the following form
\footnote{It's worth noting that we've already used the familiar form for the material fields $\mathcal{L}_m$ 
by the rescaling $B_{\mu \nu }\rightarrow \frac{qB_{\mu \nu }}{L}$ and $A_{\mu}\rightarrow \frac{qA_{\mu \nu}}{L}$.}
\cite{Chen:2010mk}.
\begin{eqnarray}
	S&=&\int d^4x\sqrt{-g}\left[\frac{1}{2\kappa^2} \left(R-2\varLambda  \right)
	-\frac{1}{4}F_{\mu \nu }F^{\mu \nu }-(D_{\mu }B_{\nu \gamma })^{*}D^{\mu }B^{\nu \gamma }-m^2B_{\mu \nu }^{*}B^{\mu \nu }  \right], \label{action}
\end{eqnarray}	
where $\kappa ^2=8\pi G_N$ is the gravitational coupling. $R$ is the Ricci scalar and
$\varLambda $ is a negative cosmological constant $(\varLambda =-3/L^2)$ with the AdS radius $L$.
$F_{\mu \nu }=\triangledown_{\mu}A_{\nu}-\triangledown_{\nu }A_{\mu  }$ is the Maxwell field strength.
$B_{\mu \nu }$ is a symmetric traceless tensor with the change $q$ and mass $m$, respectively.
$D_{\mu }=\triangledown_\mu -iqA_{\mu }$ is the covariant derivative in curved spacetime.
We consider a planar Schwarzchild-AdS ansatz to study the fully back-reacted holographic superconductor. 
\begin{eqnarray}
	ds^2=-f(r)e^{-\chi (r)}dt^2+\frac{dr^2}{f(r)}+\frac{r^2}{L^2}\left(dx^2+dy^2   \right) .\label{metric1}
\end{eqnarray}
The Hawking temperature of this black hole \eqref{metric1}, which will be interpreted as
the temperature of the dual CFT, is given by
\begin{eqnarray}
	T=\frac{e^{-\chi/2}f^{'}}{4\pi }\Bigg| _{r=r_{+}} \label{HT} .
\end{eqnarray}
For the convenience of calculation, 
we change the above metric into the following form after the coordinate transformation $r=r_{+}/z$.
In the rest of the calculation, we will set $r_{+}=1$.
\footnote{This is allowed because we have the scaling symmetry 
$r\rightarrow br, (t,x,y)\rightarrow b^{-1}(t,x,y), f\rightarrow b^2f, \phi \rightarrow b\phi$.}.
At this time, $z=0$ is the boundary where the dual d-wave superconductor lives,
\begin{eqnarray}
	ds^2=-f(z)e^{-\chi (z)}dt^2+\frac{dz^2}{z^4f(z)}+\frac{1}{z^2L^2}\left(dx^2+dy^2   \right). \label{metric2}
\end{eqnarray}
The d-wave superconductor condensate on the $x-y$ plane of the boundary with translational invariance. 
The condensate changes its sign under every $\pi /2$  rotation on the $x-y$ plane 
and the rotational symmetry is broken down to $Z(2)$.
Considering these characteristics and simplifying the equations of motion as much as possible,
we choose two ansatzes that the spatial components of symmetric traceless tensor field are 
$B_{\mu \nu }$: $B_{xx}=-B_{yy}=\psi(z)/(\sqrt{2}z^2) $ or $B_{xy}=B_{yx}=\psi(z)/(\sqrt{2}z^2)$.  
If either of the two ansatzes is chosen, the other must vanish.
Two ansatzes are equivalent under a $\theta=\pi/4$ rotation on the $x-y$ plane, 
and the equations of motion for both ansatzes are the same
\begin{eqnarray}
	{\begin{pmatrix}  
		B_{xx}  \\  
		B_{xy}
	\end{pmatrix}}\rightarrow 
	{\begin{pmatrix}  
		\cos{2\theta}&-\sin{2\theta} \\  
		\sin{2\theta}&\cos{2\theta }
	\end{pmatrix}}{\begin{pmatrix}  
		B_{xx}  \\  
		B_{xy}
	\end{pmatrix}} .
\end{eqnarray}
We consider one of them for the tensor field $B_{\mu \nu }$ and gauge field $A_{\mu }$, i.e, 
\begin{eqnarray}
	B_{xx}=-B_{yy}=\frac{\psi(z)}{\sqrt{2}z^2},\     \   A_{\mu }dx^{\mu }=\phi(z)dt \label{ansatz}.
\end{eqnarray}	
After the variation of the action \eqref{action},	
the equations of motion can be easily obtained
\footnote{Notice that we have hidden the AdS radial $L$ in the above equations of motion, 
since we have taken $L=1$.
We set $L=1$ because we have the scaling symmetries 
$L\rightarrow aL,\   \ z\rightarrow a^{-1}z, \   \ t\rightarrow at,\   \ q\rightarrow a^{-1}q$.}
\begin{eqnarray}
	&&\psi^{''}(z)+\psi^{'}(z)\left[\frac{f^{'}(z)}{f(z)}-\frac{\chi ^{'}(z)}{2}    \right]+\psi (z)\left[\frac{q^2e^{\chi (z)}\phi (z)^2}{z^4f(z)^2}-\frac{m^2}{z^4f(z)}-\frac{2}{z^2}   \right]=0 \label{eq1}   ,\\
	&&\phi^{''}(z)+\frac{\chi^{'} (z)\phi^{'}(z)}{2}-\frac{2q^2\phi(z)\psi(z)^2}{z^4f(z)}=0 \label{eq2}  ,\\
	&&\chi ^{'}(z)-2\kappa ^2z\left\{\psi^{'}(z)^2+\psi(z)^2\left[\frac{2}{z^2}+\frac{q^2e^{\chi (z)}\phi (z)^2}{z^4f(z)^2}    \right]    \right\}=0\label{eq3},\\
	&&f^{'}(z)-\frac{f(z)}{z}+\frac{3}{z^3}-\kappa^2 z \left\{\frac{e^{\chi (z)}\phi ^{'}(z)^2}{2}+f(z)\psi^{'}(z)^2+\psi(z)^2 \left[\frac{m^2}{z^4}+\frac{q^2e^{\chi (z)}\phi (z)^2}{z^4f(z)}+\frac{2f(z)}{z^2} \right]    \right\}=0.   \nonumber\\
	&&       \label{eq4}
\end{eqnarray}	
Without loss of generality, we will set $q=1$ and keep $\kappa ^2$ finite when we consider the backreactions
\footnote{If one re-scales the tensor field $B_{\mu \nu }$ 
and the gauge potential $A_{\mu }$ to $B_{\mu \nu }\rightarrow \frac{B_{\mu \nu }}{q}$ and $A_{\mu}\rightarrow \frac{A_{\mu }}{q}$ in the action, 
the tensor field and Maxwell field equations \eqref{eq1}$-$\eqref{eq2} will remain invariant, 
but the gravitational coupling coefficient $\kappa^2 $ of the Einstein field equations 
\eqref{eq3}$-$\eqref{eq4} is re-scalared by $\kappa^2\rightarrow \frac{\kappa^2}{q^2}$.
One can fix the charge density $q=1$ and vary $\kappa$ which is used to reflect
the strength of the backreaction.}.
Notice that these equations of motion are very similar to those for s-wave superconductors \cite{Liu:2011fy,Pan:2012jf}.
We will numerically solve the above equations by using the shooting method.
Firstly, we need to know not only the behavior of the equations of motion at the horizon
but also the behavior of those equations toward the asymptotical AdS boundary.
Because of $f(z_h)=0$ and $\phi(z_h)=0$
\footnote{To keep the norm of the gauge potential $A_{\mu}A^{\mu}=g^{tt}\phi^2$ 
finite at the horizon, we must make the gauge potential $\phi $ vanish at the horizon.}
those fields equations \eqref{eq1}$-$\eqref{eq4} can be expanded at the horizon $(z_{h}=1)$,
\begin{eqnarray}
	&&\psi(z)=\psi_{0}+\psi_{1}(1-z)+\cdot \cdot \cdot,\    \ \phi(z)=\phi_{1}(1-z)+\cdot \cdot \cdot,    \\  \label{ex1}
	&&f(z)=f_{1}(1-z)+\cdot \cdot \cdot,\     \  \chi (z)=\chi_{0}+\chi_{1}(1-z)+\cdot \cdot \cdot,  \nonumber
\end{eqnarray}
where the coefficients $\psi_0$, $\phi_{1}$, and $\chi_0$ are some constants.
To obtain $\psi(z)$, $\phi(z)$, $\chi(z)$, and $f(z)$ that satisfy the asymptotic boundary conditions of the equations,
we can solve the equations of motion \eqref{eq1}$-$\eqref{eq4} by performing the shooting method.
Before that, we need to know the asymptotic behavior of the equations of motion near the AdS boundary $(z\rightarrow 0)$,
\begin{eqnarray}
	\chi(z)\approx \chi_{c} ,\     \ f(z)\approx \frac{1}{z^2},\   \  \phi\approx  \mu -\rho z,\     \  \psi\approx  z^{\Delta _{-}}\psi _{-}+z^{\Delta _{+}}\psi_{+},
\end{eqnarray}
where $\Delta _{\pm }=\frac{3 \pm \sqrt{17+4m^2}}{2}$ with the BF bound $m^2\geq -17/4$. 
If $\Delta_{-}\leq 0$  $(m^2\geq -2)$, the normalizability requires that $\psi_{-}$ must vanish.
According to the gauge/gravity dictionary, the coefficient $\psi_{+}$ corresponds to the vacuum expectation value of the operator 
dual to the components of the tensor field $B_{\mu \nu }$, and
$\psi_{-}$ is dual to a source for this operator.
In this paper, we consider $m^2=-1/4$ (i.e. $\Delta_{-}=-1/2$ and $\Delta_{+}=7/2$).
To get the metric \eqref{metric2} back to the standard AdS metric at the boundary, let us set the constant $\chi_{c}$ equal to zero $\chi_{c}=0$
\footnote{This is allowed by the rescaling symmetry 
$e^{\chi}\rightarrow a^2e^{\chi},\   \ \phi\rightarrow \phi /a, \    \ t\rightarrow at $ for a particular value $a$.}.
Here $\mu$ is interpreted as the chemical potential and $\rho $ as the charge density in the boundary theory. 
The condensate of the tensor operator $\left\langle \mathcal{O}_{ij}\right\rangle $ 
in the boundary field theory dual to the field $B_{\mu \nu }$ is given by
\begin{equation}
\left\langle \mathcal{O}_{ij}\right\rangle =\left(\begin{matrix}
	\psi_{+}&0\\
	0& -\psi_{+}
\end{matrix}\right).
\end{equation}

\begin{figure}[htb]
	\centering
	\subfigure[]{
	\includegraphics[width=0.475\textwidth,height=0.35\textwidth]{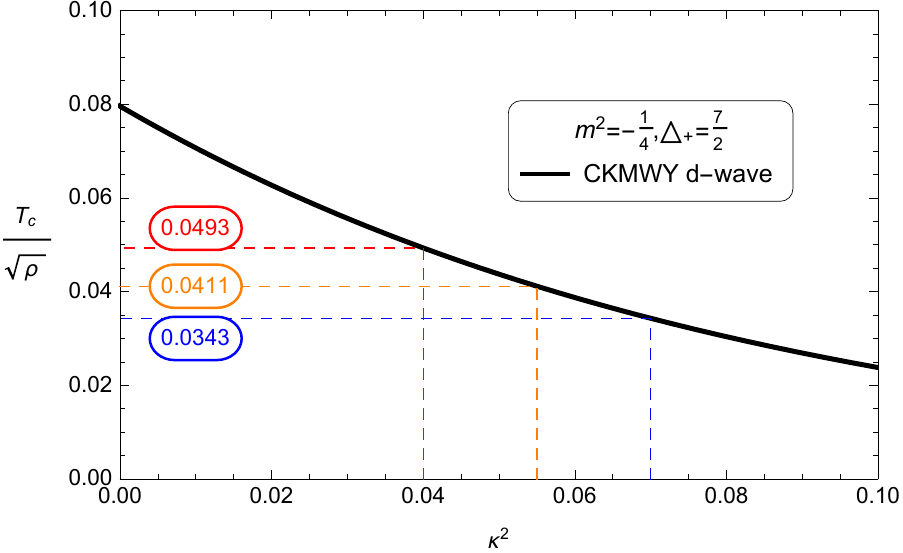}}
    \hfill
	\subfigure[]{
	\includegraphics[width=0.475\textwidth,height=0.35\textwidth]{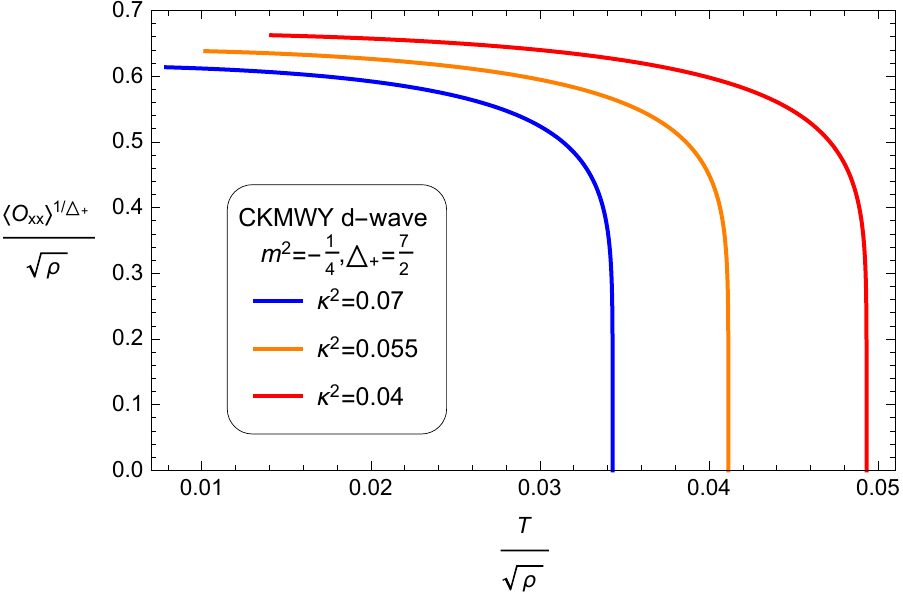}}
	\caption{The left figure depicts the change of critical temperature with backreaction $\kappa^2$
	for $m^2=-1/4$ in the CKMWY d-wave superconducting model. The right figure is the condensate as a function of temperature.
	Different color solid lines correspond to different backreaction values, 
	the critical temperatures are 
	$T_{c}/\sqrt{\rho }=0.0343$  (blue: $\kappa^2=0.07$),
	$T_{c}/\sqrt{\rho }=0.0411$  (orange: $\kappa^2=0.055$) and
	$T_{c}/\sqrt{\rho }=0.0493$   (red: $\kappa^2=0.04$) respectively. } \label{ckmwy}
\end{figure}

In Fig. \ref{ckmwy}a, we plotted the critical temperature of the superconducting phase transition as a function of the backreactions.
It was easy to see that the critical temperature decreases monotonically with the increase of the backreactions.
Our results show that the stronger backreactions will also make the operator condense more difficult in d-wave superconductors.
In Fig. \ref{ckmwy}b, the condensations of the operator $\left\langle \mathcal{O}_{xx} \right\rangle$ for some backreactions were investigated.
One can fit these curves near the critical point and observe that the expected value of the operator behaves like 
$\left\langle \mathcal{O}_{xx} \right\rangle \sim \left(T_{c}-T\right)^{1/2}$,
which implies that the d-wave superconducting phase transition is a second-order phase transition with the exponent $1/2$.

Finally, it is essential to emphasize that the physical quantities we want to obtain are independent of the scaling scale.
Notice that, according to the scaling symmetry 
\begin{eqnarray}
	(t,z,x,y)\rightarrow b^{-1}(t,z,x,y),\   \ f\rightarrow b^2f,\    \  \phi \rightarrow b \phi, \label{scaling}
\end{eqnarray}	
the quantities $T,\mu ,\rho$ and $\left\langle \mathcal{O}_{xx}\right\rangle $ scale as 
\begin{eqnarray}
	T\rightarrow b T,\   \  \mu \rightarrow b\mu,\    \ \rho \rightarrow b^2 \rho,\    \  \left\langle\mathcal{O}_{xx}\right\rangle\equiv\psi_{+}\rightarrow b^{\bigtriangleup_{+}}\psi_{+}.
\end{eqnarray}
Therefore, to analyse the physics, it is helpful to make these quantities dimensionless, i.e.,
$T/\sqrt{\rho }$, $\left\langle \mathcal{O}_{xx}\right\rangle^{1/\bigtriangleup_{+}}/\sqrt{\rho } $.

\subsection{BHRY D-wave superconductor}

With the same strategy, in this section, 
we study the effect of backreactions on the d-wave model proposed by BHRY.
This holographic model with a symmetric tensor field $\varphi _{\mu \nu }$ 
and a U(1) gauge field $A_{\mu }$ is described by the following action \cite{Benini:2010pr}
\begin{eqnarray}
	S&=&\int d^4x\sqrt{-g}  \left\{\frac{1}{2\kappa^2} \left(R+\frac{6}{L^2}\right) +\mathcal{L}_m \right\} ,
\end{eqnarray}	
with
\begin{eqnarray}
	\mathcal{L}_m&=&-\frac{1}{4}F_{\mu \nu }F^{\mu \nu }-|D_{\rho}\varphi _{\mu \nu }|^2
	+2|D_{\mu} \varphi ^{\mu \nu }|^2+|D_{\mu }\varphi |^2-\left(D_{\mu }\varphi ^{\mu \nu }D_{\nu}\varphi +c.c.\right)
	 -m^2(|\varphi _{\mu \nu }|^2-|\varphi |^2  )\nonumber\\
	 &&+2R_{\mu \nu \rho \lambda }\varphi ^{\mu \rho }\varphi ^{\nu \lambda } -\frac{1}{4}R|\varphi|^2-iqF_{\mu \nu }
	 \varphi ^{\mu \lambda }\varphi^{\nu }_{\lambda } ,\nonumber                   \label{bhrya1}
\end{eqnarray}	
where $D_{\mu }=\triangledown_{\mu }-iqA_{\mu }$, $\varphi\equiv \varphi ^{\mu }_{\mu }$, 
and $R^{\mu }_{\nu \rho \lambda }$ is the Riemann tensor of the background metric.
The parameter $q$ is the charge of the massive spin-two fields.
We shall set $q=1$ without loss of generality in the following discussion.
An ansatz where $\varphi _{\mu \nu }$ and $A_{\mu }$ depend only on the radial coordinate $z$ is considered,  
and only the two spatial components of $\varphi $ are turned on
\begin{eqnarray}
	\varphi_{xx}(z)=-\varphi_{yy}(z)=\frac{\psi(z)}{\sqrt{2} z^2},  \    \   A_{\mu }dx^{\mu}=\phi (z)dt,  \label{ans}
\end{eqnarray}
with all other components of $\varphi_{\mu \nu }$ setting to zero, and the real $\phi $ and $\psi$.
The ansatz \eqref{ans} satisfies $\varphi =\varphi _{\mu }=F_{\mu \rho }\varphi ^{\rho }_{\nu }=0$.
There are only four nonzero terms left in the material action $\mathcal{L}_m$
\footnote{After changing $\varphi_{\mu \nu }$ to $B_{\mu \nu }$, 
the matter part $\mathcal{L}_{m}$ of the action for this model is very similar to the one for the CKMWY d-wave model 
in the previous section except for the last term $2R_{\mu \nu \rho \lambda }\varphi ^{*\mu \rho }\varphi ^{\nu \lambda }$.}

\begin{eqnarray}
	\mathcal{L}_m=-\frac{1}{4}F_{\mu \nu }F^{\mu \nu } -|D_{\rho}\varphi _{\mu \nu }|^2-m^2|\varphi _{\mu \nu }|^2
	+2R_{\mu \nu \rho \lambda }\varphi ^{*\mu \rho }\varphi ^{\nu \lambda }. \label{bhrya2}
\end{eqnarray}
Using the planar Schwarzchild-AdS ansatz \eqref{metric2}, we can quickly get the equations of motion.
\footnote{We've set the AdS radius $L=1$ here. }
\begin{eqnarray}
	&&\psi^{''}(z)+\psi^{'}(z)\left[\frac{f^{'}(z)}{f(z)}-\frac{\chi ^{'}(z)}{2}    \right]+\psi (z)\left[\frac{q^2e^{\chi (z)}\phi (z)^2}{z^4f(z)^2}-\frac{m^2}{z^4f(z)}  \right]=0 ,\label{eq5}\\
	&&\phi^{''}(z)+\frac{\chi^{'} (z)\phi^{'}(z)}{2}-\frac{2q^2\phi(z)\psi(z)^2}{z^4f(z)}=0 ,\label{eq6}\\
	&&\chi ^{'}(z)-2\kappa ^2z\left[\psi'(z)^2+\frac{q^2e^{\chi(z)}\phi(z)^2\psi(z)^2}{z^4f(z)^2} \right] =0 , \label{eq8}\\
	&&f^{'}(z)-\frac{f(z)}{z}+\frac{3}{z^3}-\kappa ^2z\left\{\frac{e^{\chi(z)}\phi'(z)^2}{2}+f(z)\psi'(z)^2+\frac{\psi(z)^2}{z^4}\left[m^2+\frac{q^2e^{\chi(z)}\phi(z)^2}{f(z)}\right]  \right\}=0  .\nonumber\\
	&&                           
\end{eqnarray}	
The tensor field's charge was set to $q = 1$ without losing generality.
We can solve these equations of motion by the shooting method, which detail has been introduced in the previous section.
Near the asymptotically AdS boundary, the Maxwell field $\phi$, and tensor field $\psi$ behaves as 
\begin{eqnarray}
	\phi\approx  \mu -\rho z,\     \  \psi\approx  z^{\Delta _{-}}\psi _{-}+z^{\Delta _{+}}\psi_{+},
\end{eqnarray}
where $\Delta _{\pm }=\frac{3 \pm \sqrt{9+4m^2}}{2}$ with the stability bound $m^2\geq 0$. 
To keep the conformal dimension $\Delta_+$ the same as that of the CKMWY d-wave model studied in the previous subsection,
we would like to consider the mass of tensor field $m^2=7/4$, i.e., $\Delta_{-}=-1/2$ and $\Delta_{+}=7/2$.
According to the gauge/gravity dual dictionary, $\left\langle \mathcal{O}_{xx}\right\rangle\equiv \psi_{+}$ 
is interpreted as the expected value of the operator $\mathcal{O}_{xx}$, and 
$\psi_{-}$ is dual to a source for this operator and must vanish.
We can also read out the chemical potential $\mu $ and the related charge density $\rho $ from the expansion of $\phi (z)$.

\begin{figure}[htb]
	\centering
	\subfigure[]{
	\includegraphics[width=0.475\textwidth,height=0.35\textwidth]{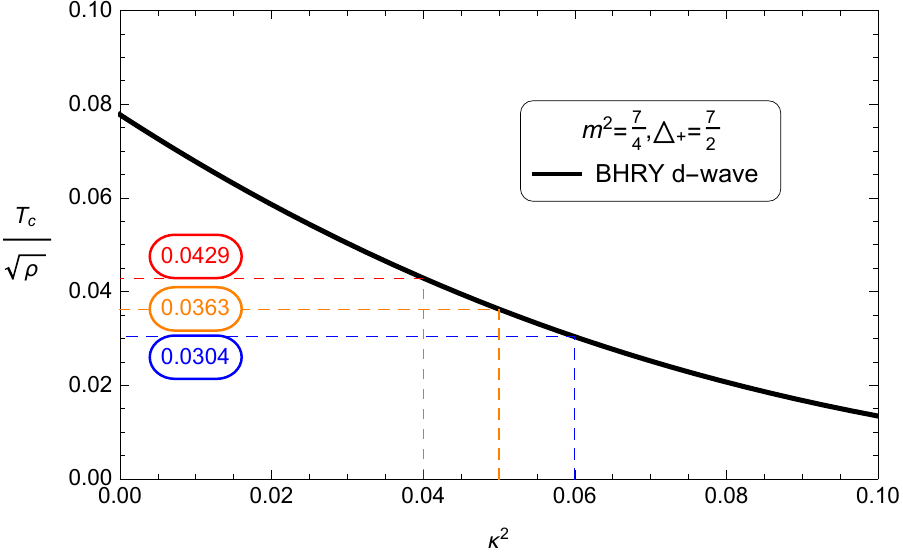}}
    \hfill
	\subfigure[]{
	\includegraphics[width=0.475\textwidth,height=0.35\textwidth]{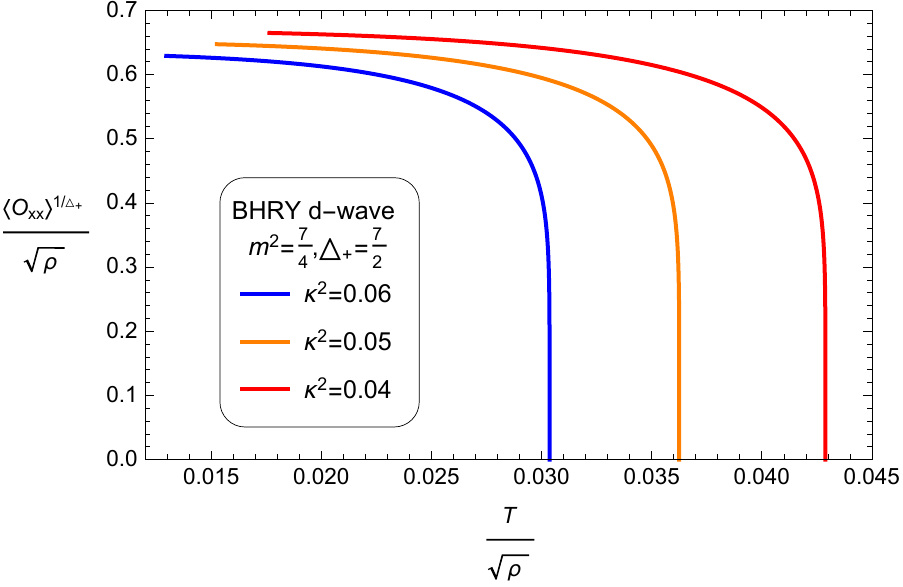}}
	\caption{The left figure depicts the change of critical temperature with backreaction $\kappa^2$
	for $m^2=7/4$ in the BHRY d-wave superconducting model.
	The right figure is the condensate as a function of temperature.
	Different colors correspond to different backreaction values, 
	and the critical temperatures are 
	$T_{c}/\sqrt{\rho }=0.0304$     (blue:   $\kappa^2=0.06$),
	$T_{c}/\sqrt{\rho }=0.0363$     (orange:   $\kappa^2=0.05$) and
	$T_{c}/\sqrt{\rho }=0.0429$     (Red:    $\kappa^2=0.04$)  respectively. 
	The condensate goes to zero at $T=T_{c}\propto \sqrt{\rho } $.} \label{bhry}
\end{figure}

In Fig. \ref{bhry}a, we draw the critical temperature of phase transition with the strength of backreactions $\kappa^2$.
The curve in this model is similar to that in the CKMWY d-wave model.
It is interesting to note that the critical temperature of this d-wave model 
is slightly smaller than that of the CKMWY d-wave model (see the red line)
when both the parameters $\kappa$ and $\Delta_+$ between the two models keep the same.   
Therefore, we can conclude that the non-zero curvature term $2R_{\mu \nu \rho \lambda }\varphi ^{*\mu \rho }\varphi ^{\nu \lambda }$ 
in action \eqref{bhrya2} makes the superconducting phase transitions more difficult.
In Fig. \ref{bhry}b, we also investigated the condensations of the operator $\left\langle \mathcal{O}_{xx} \right\rangle$ 
for three different backreactions.
Near the critical temperature, the expected value of the operator behaves like
 $\left\langle \mathcal{O}_{xx} \right\rangle \sim \left(T_{c}-T\right)^{1/2}$, which is a second-order phase transition.

\section{HEE and HSC for D-Wave Holographic Superconductors} \label{sec3}

In this section, we will numerically study the HEE 
and HSC for a strip subregion of the 2-dimensional boundary system for the CKMWY and BHRY d-wave superconducting models.
We can get the induced metric from the linear element \eqref{metric2} when fixing the time
\begin{eqnarray}
	ds^{2}_{induced}=\left[\frac{1}{z^4f(z)}\left(\frac{dz}{dx}\right)^2+\frac{1}{z^2L^2} \right]dx^2+\frac{1}{z^2L^2}dy^2. 
\end{eqnarray}
Then the area for a strip subsystem $\mathcal{A}$ with the range $x\subseteq \left[-\frac{L_{x}}{2}, \frac{L_{x}}{2}   \right] $ and 
$y\subseteq \left[-\frac{L_{y}}{2}, \frac{L_{y}}{2}   \right] $ can be given by the following formula
\begin{eqnarray}
	Area\left( \mathcal{A}\right) =2L_{y}\int^{\frac{L_x}{2}}_{0} \frac{dx}{z^3}\sqrt{\frac{1}{f}\left(  \frac{dz}{dx}  \right)^2+z^2}.   \label{subregion}
\end{eqnarray}
The area of the minimal surface $\gamma_{\mathcal{A}}$
can be obtained by making a variation of the integrand function above,
and the HEE can be obtained by the RT formula \eqref{HEE}. 
Taking the variational of the integrand function \eqref{subregion} for $x$ 
and considering the boundary condition $\frac{dz}{dx}\vert _{z=z_{*}}=0$, 
the following integral equation can be derived
\begin{eqnarray}
	\frac{dz}{dx}=\sqrt{\frac{f(z^{4}_{*}-z^4)}{z^2}} \label{HEEeq1},
\end{eqnarray}
 where $z_{*}$ is the turning point of the minimal surface.
There is another boundary condition, which is the regular boundary of the asymptotic infinity $z\rightarrow 0$
\begin{eqnarray}
	\int^{z_{*}}_{\varepsilon \rightarrow 0} \frac{zdz}{\sqrt{f(z^4_*-z^4)}} =\frac{L_{x}}{2},
\end{eqnarray}
where $\varepsilon $ is a UV cutoff used to avoid the divergent problem 
and will be taken as a small value in our numerical strategy. 
Substituting Eq. \eqref{HEEeq1} into Eq. \eqref{subregion} and using the RT formula \eqref{HEE}. We can get the total HEE $\mathcal{S}$
\begin{eqnarray}
	\mathcal{S}=\frac{2 L_y}{4G_{N}}\int^{z_*}_{\epsilon }\frac{z^2_*dz}{z^3\sqrt{f(z^4_*-z^4)} }
	=\frac{ L_y}{2G_{N}}\left(S_{u}+\frac{1}{\varepsilon } \right),  \label{HEE2}
\end{eqnarray}
where $S_{u}$ is the universal term that is physically important. 
$1/\varepsilon $ is a divergent term that comes from pure AdS background 
and can be removed by the regularization where the computational details are included in Appendix \ref{Appendix}. 
From Eq. \eqref{HEEeq1}, 
we can also get the integral expression $x(z)$ about $z$ for a minimal surface 
\begin{eqnarray}
	x(z)=\int ^{z_*}_{z}\frac{zdz}{\sqrt{f(z^4_*-z^4)}}. \label{minsur}
\end{eqnarray}
The volume enclosed by the minimal surface $\gamma_{A}$ and the strip region $\mathcal{A}$ 
can be obtained from the volume integration of the minimal surfaces in bulk
\begin{eqnarray}
	V(\gamma_{A})=2L_{y}\int^{z_*}_{\epsilon }\frac{x(z)dz}{z^4\sqrt{f} }.
\end{eqnarray}
Using  Eq.\eqref{HSC}, we can obtain the total HSC
\begin{eqnarray}
	\mathcal{C}=\frac{2L_{y}}{8\pi L G_{N}}\int^{z_*}_{\epsilon }\frac{x(z)dz}{z^4\sqrt{f} }
	=\frac{L_{y}}{4\pi LG_N}\left[C_{u}+\frac{\mathcal{F}(z_{*})}{\varepsilon^2}   \right] ,\label{HSC2}
\end{eqnarray}
where $C_{u}$ is the universal term, and $\mathcal{F}(z_{*})/\varepsilon^2$ is the diverging term.
The divergence term comes from the pure AdS background where the computational details are included in Appendix \ref{Appendix}. 
Although the divergent term does not have an exact expression, we can still solve it numerically.
Since the finite term does not change with the cutoff $\varepsilon $, 
we can subtract the complexity from each other that corresponds to two different cutoffs $\varepsilon_{1}$ and $\varepsilon_{2}$,
so that the finite term can be eliminated and $\mathcal{F}(z_{*})/(\varepsilon^2_1-\varepsilon^2_2)$ will be left, 
where the numerator $\mathcal{F}(z_*)$ of the divergent term will be obtained numerically.
Under the scaling symmetries of Eq. \eqref{scaling}, we can rescale the $L_{x}$, $S_{u}$, and $C_{u}$ as  
\begin{eqnarray}
	L_{x}\rightarrow b^{-1}L_{x}, \   \  S_{u}\rightarrow bS_{u},\    \  C_{u}\rightarrow bC_{u}.
\end{eqnarray}
Therefore, the dimensionless forms of these quantities are helpful
\begin{eqnarray}
	L_{x}\sqrt{\rho },  \    \ \frac{S_{u}}{\sqrt{\rho } }, \    \   \frac{C_{u}}{\sqrt{\rho } }.
\end{eqnarray}

\begin{figure}[htb]
	\centering
	\subfigure[]{
	\includegraphics[width=0.475\textwidth,height=0.35\textwidth]{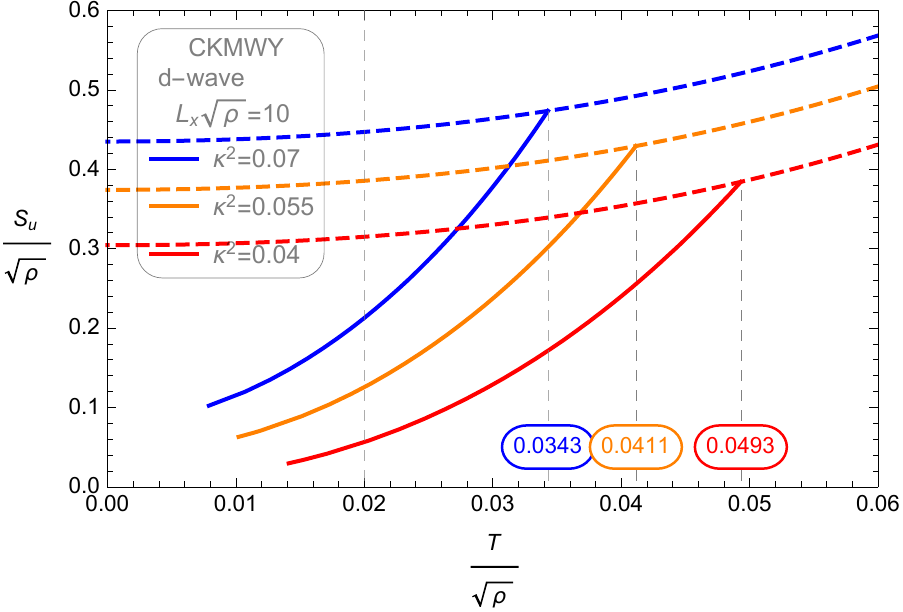}}
    \hfill
	\subfigure[]{
	\includegraphics[width=0.475\textwidth,height=0.35\textwidth]{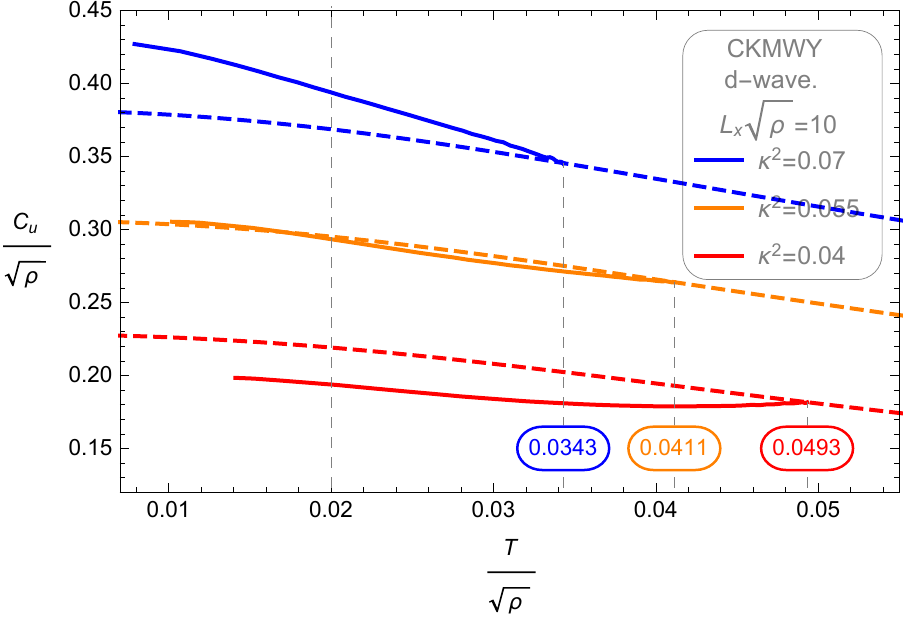}}
	\caption{The HEE (a) and the HSC (b) as functions of
	 the temperature $T/\sqrt{\rho } $ for the CKMWY d-wave superconductor with a fixed subregion $L_{x}$: $L_{x}\sqrt{\rho }=10$.
	The dashed and solid curves correspond respectively to normal and superconducting phases. } \label{ckmwyheehsc}
\end{figure}
\begin{figure}[htb]
	\centering
	\subfigure[]{
	\includegraphics[width=0.475\textwidth,height=0.35\textwidth]{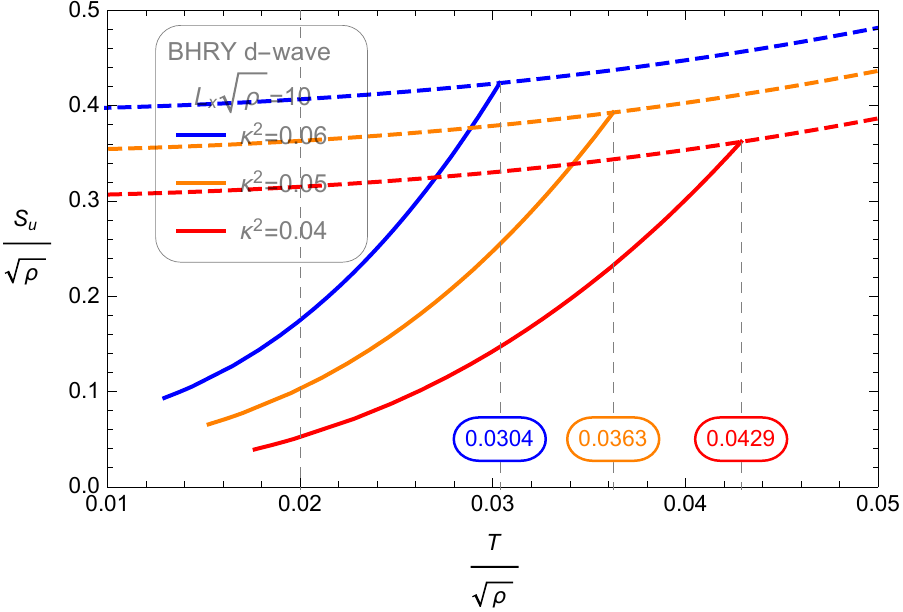}}
    \hfill
	\subfigure[]{
	\includegraphics[width=0.475\textwidth,height=0.35\textwidth]{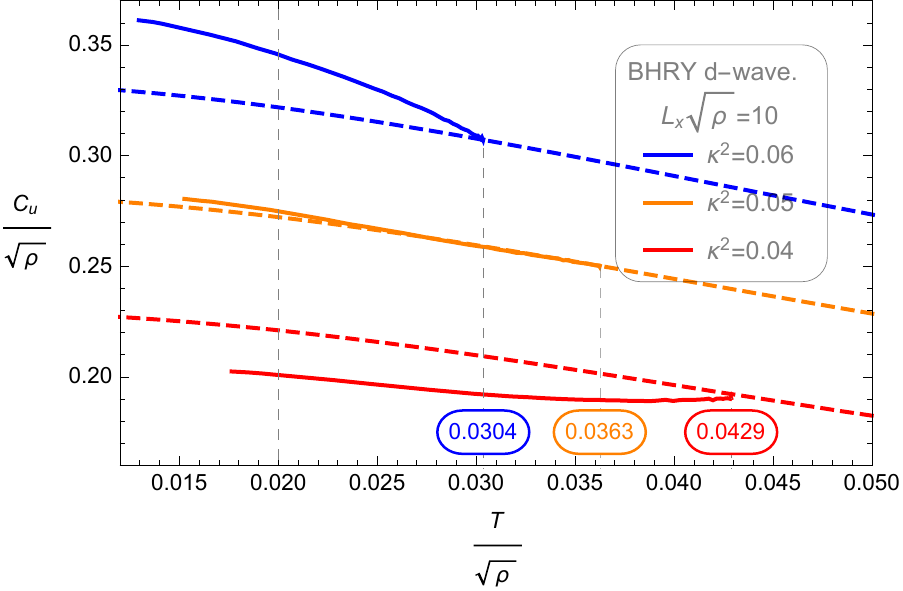}}
		\caption{The HEE (a) and the HSC (b) as functions of
		 the temperature $T/\sqrt{\rho } $ for the BHRY d-wave superconductor with a fixed subregion $L_{x}$: $L_{x}\sqrt{\rho }=10$.
		The dashed and solid curves correspond respectively to normal and superconducting phases.  } \label{bhryheehsc}
\end{figure}

In Figs. \ref{ckmwyheehsc} and \ref{bhryheehsc}, 
we plot the HEE and HSC as functions of temperature $T/\sqrt{\rho } $ 
for two holographic d-wave superconducting models 
when the strip-width of subregion $L_{x}$ is fixed.
From sub-figure $(a)$ of Figs. \ref{ckmwyheehsc} and \ref{bhryheehsc},
we notice that the HEE decreases gradually as the temperature decreases, and
finally approaches a non-zero constant at zero temperature, just as shown in \cite{Albash:2012pd,Chakraborty:2019vld} 
for the HEE of the operator in the holographical s-wave superconductor.
In addition, we also find that the HEE increases with the increase of the backreactions $\kappa^2$ for a fixed temperature. 
When considering the effects of non-zero backreactions,
we can see that the HEE of the superconducting phase is always lower than 
that of the normal phase for a given temperature $T$ below the critical point.
As the degrees of freedom condense below the critical temperature, the entropy exhibits a lower value,
similar to the result of s-wave superconductors \cite{KordZangeneh:2017zyy,Chakraborty:2019vld}.
It is also noted that there is a slope discontinuity at the point of critical temperature $T_{c}$, which 
suggests that the HEE can also be used as an independent probe for d-wave superconducting phase transitions 
not just in s-wave and p-wave superconductors.
We note that the behaviors of the HEE for these two d-wave superconducting models are very similar, 
except for the difference of transformation temperature.

\begin{figure}[htb]
	\centering
	\subfigure[]{
	\includegraphics[width=0.475\textwidth,height=0.35\textwidth]{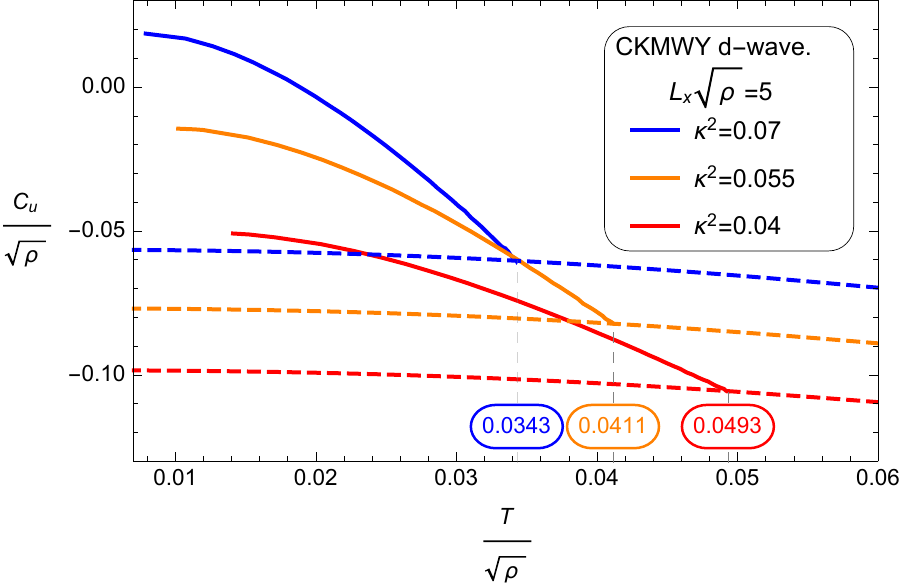}}
    \hfill
	\subfigure[]{
	\includegraphics[width=0.475\textwidth,height=0.35\textwidth]{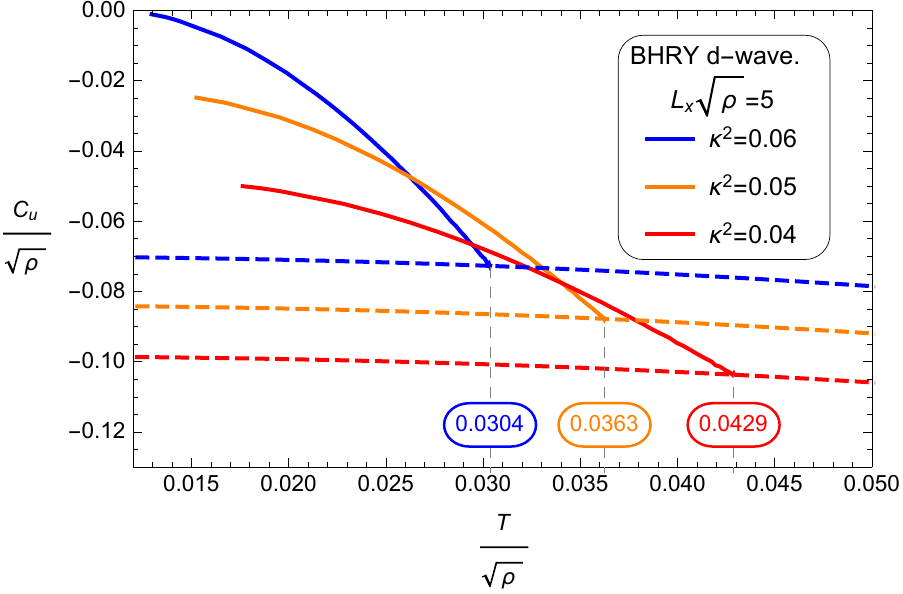}}
	\caption{The HSC as functions of the temperature $T/\sqrt{\rho } $ for the CKMWY (a) 
	and the BHRY (b) d-wave superconductor with a fixed subregion $L_{x}$: $L_{x}\sqrt{\rho }=5$.
	The dashed and solid curves correspond respectively to normal and superconducting phases. } \label{ckbhhsc03}
\end{figure}

\begin{figure}[htbp]
	\centering
	\subfigure[]{
	\includegraphics[width=0.475\textwidth,height=0.35\textwidth]{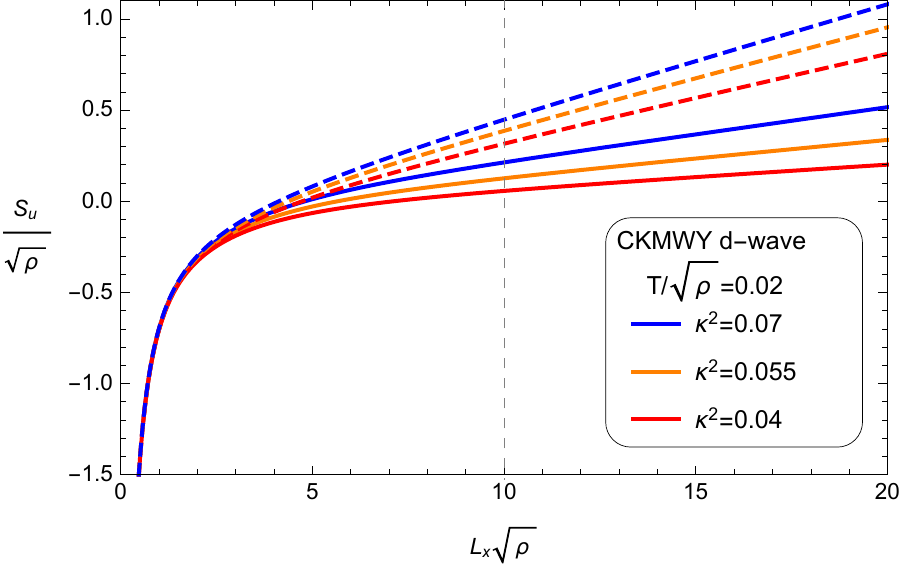}}
    \hfill
	\subfigure[]{
	\includegraphics[width=0.475\textwidth,height=0.35\textwidth]{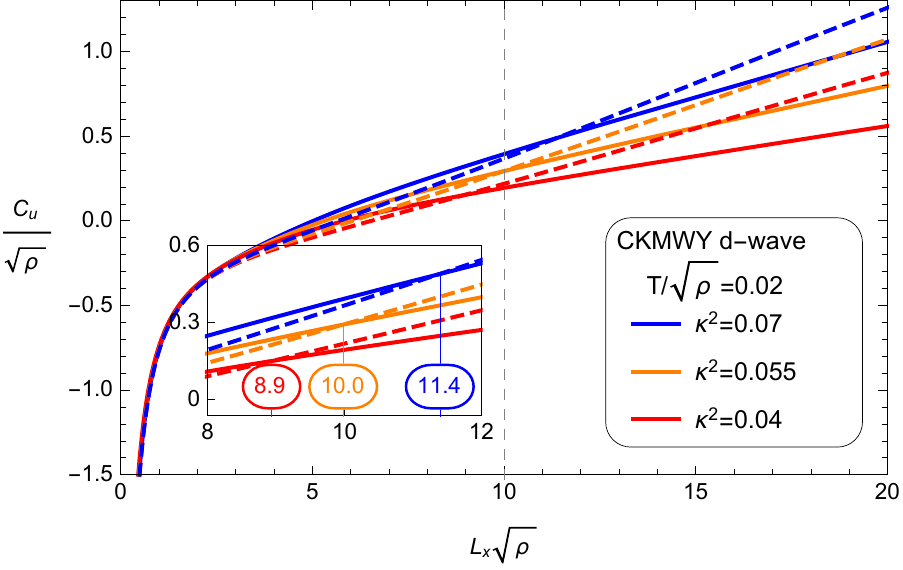}}
			\caption{The HEE (a) and the HSC (b) as functions of 
			the strip-width $L_{x}\sqrt{\rho } $ for the CKMWY d-wave superconductor with a fixed temperature $T$: $T/\sqrt{\rho } =0.02$.
			The dashed and solid curves correspond respectively to normal and superconducting phases.  } \label{ckmwyheehsc2}
\end{figure}

\begin{figure}[htbp]
	\centering
	\subfigure[]{
	\includegraphics[width=0.475\textwidth,height=0.35\textwidth]{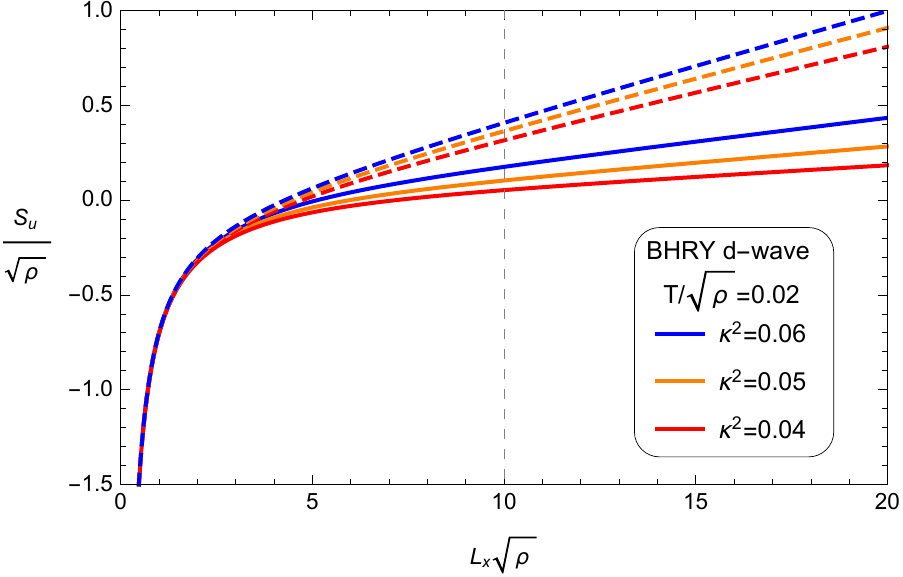}}
    \hfill
	\subfigure[]{
	\includegraphics[width=0.475\textwidth,height=0.35\textwidth]{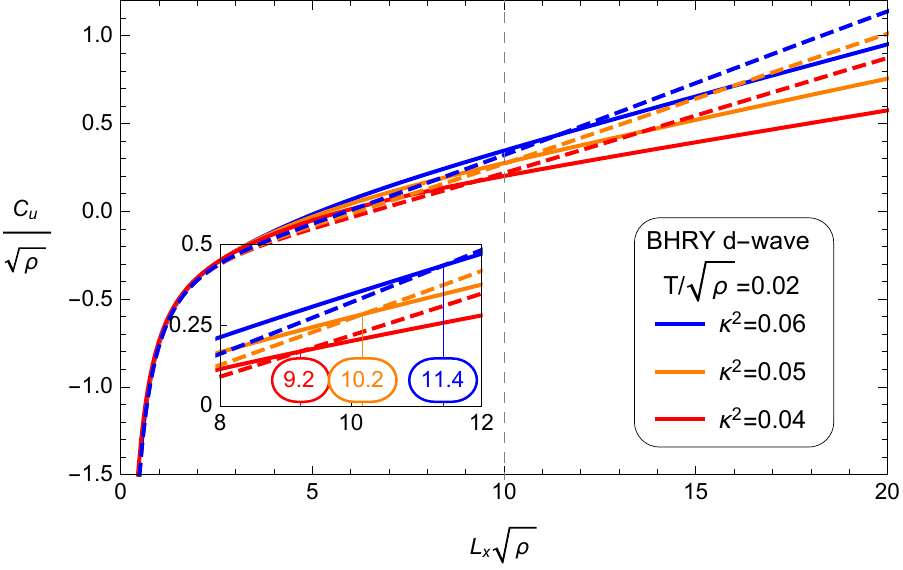}}
				\caption{The HEE (a) and the HSC (b) as functions of 
				the strip-width $L_{x}\sqrt{\rho }$ for the BHRY d-wave superconductor with a fixed temperature $T$: $T/\sqrt{\rho } =0.02$.
				The dashed and solid curves correspond respectively to normal and superconducting phases.  } \label{bhryheehsc2}
\end{figure}

The HSC of the two d-wave superconducting models is shown separately 
in the second subfigure $(b)$ of Figs. \ref{ckmwyheehsc} and \ref{bhryheehsc}.
We find that the slope of the HSC with temperature is discontinuous at the critical point $T_{c}$,
which indicate that the HSC can also be used as probe for the critical temperature $T_{c}$ 
of the d-wave superconducting phase transition.
Different from the behavior of the HEE, the behaviors of the HSC are more complicated. 
Specifically, we observe that the behaviors of the HSC for d-wave superconductors
depend on both the strip-width $L_{x}\sqrt{\rho }$ of the subregion and the backreactions $\kappa $.
When $L_{x}\sqrt{\rho } $ takes some values (like $L_{x}\sqrt{\rho }=5$ in Fig. \ref{ckbhhsc03})
smaller than the critical value or the backreactions $\kappa^2$ larger than the critical value
(these critical values of the strip-width are given in the subfigure (b) of Figs. \ref{ckmwyheehsc2}, \ref{bhryheehsc2}), 
the superconducting phase always has a larger complexity than the normal phase below the critical temperature.
At this time, the HSC $C_{u}$ for the superconducting phase decreases with the increase of temperature $T/\sqrt{\rho }$, 
which is similar to the result in Ref \cite{KordZangeneh:2017zyy}.
When $L_{x}\sqrt{\rho } $ takes some values (such as $L_{x}\sqrt{\rho }=10$ in our Figs. \ref{ckmwyheehsc}, \ref{bhryheehsc})
larger than the critical value and the backreactions $\kappa^2$ smaller 
than the critical value (such as the red line with $\kappa^2=0.04$ in our Figs. \ref{ckmwyheehsc}, \ref{bhryheehsc}),
we find that the HSC first decreases and then increases slightly as we lower the temperature 
during the superconducting phase, 
which is similar to the result of $(2+1)$-dimensional holographic $\mathcal{O}_{1}$ order superconductor 
in Ref. \cite{Chakraborty:2019vld}.
Similar phenomena that the different behaves of the HSC for different fixed strip-widths 
in superconducting phase transitions have also been found recently in Refs. \cite{Shi:2021rnd,Shi:2020nwx}, 
but the physical mechanism is still unknown.

Furthermore, in the Figs. \ref{ckmwyheehsc2} and \ref{bhryheehsc2},
we also investigate the HEE and HSC as a function of the strip-width $L_{x}\sqrt{\rho } $ 
with the fixed temperature  $T/\sqrt{\rho }=0.02$.
We find that both HEE and HSC increase linearly with the increase of strip-width for larger values of $L_{x}\sqrt{\rho}$. 
This is known as the ``area law".  
As we mentioned, we can intuitively see  
that there is a crossing point between the HSC of the superconducting phase and that of the normal phase 
in the subfigures \ref{ckmwyheehsc2}b and \ref{bhryheehsc2}b.
The behaviors of HEE and HSC for these two d-wave superconducting models are very similar 
except for the difference in transformation temperature.


\section{ Summary}   \label{sec4}
In this paper, we investigated the HEE and HSC of two holographic d-wave superconducting models with backreactions.
Firstly, we studied the condensation characteristics of these two d-wave models after phase transition. 
Using the same ansatz and keeping the same dimensions of operators, 
we obtained the critical temperatures of phase transitions under different backreactions,
showing that the critical temperature decreases as the backreaction increases.
This indicates that the backreaction will hinder the d-wave condensation to be formed. 
It's also worth noting that, 
when keeping both the parameters $\kappa$ and $\Delta_+$ corresponding to two models the same,
the critical temperature of the BHRY model 
is slightly smaller than that of the CKMWY model.
Further analysis revealed that this slight but essential difference is due to the coupling term of the
tensor field and curvature in action \ref{ans}.
Secondly, we studied the HEE and HSC for these two holographical d-wave superconducting models.
When the strip-widths $L_{x}$ of subregions were fixed, 
we found that the slope of HEE and HSC varies with temperature and 
is discontinuous at the point of critical temperature, 
which means that both the HEE and HSC can be used as good probes 
to the phase transition in holographic d-wave superconductors.
Furthermore, when the temperature was fixed, we observed that the HEE and HSC showed similar 
linear growth behavior for large widths of subregions.
We also noted that the d-wave superconducting phase always has a lower HEE than the normal phase
due to the condensation of degrees of freedom below the critical temperature.
However, the behaviors of HSC are more complicated.
More specifically, we found that the behaviors of HSC depend on the strip-width of the subregion and the backreaction.
When the width of the subregion is smaller than a critical value or the backreaction is larger than a critical value,
the HSC of the superconducting phase is larger than that of the normal phase.
In this case, the HSC decreases with the temperature increase,
which is similar to the result in Ref. \cite{KordZangeneh:2017zyy}.
However, suppose the width of the subregion is larger than the critical value 
and the backreaction is smaller than the critical value. In this case, 
the HSC of the superconducting phase is smaller than that of the normal phase. 
The HSC first decreases and then slightly increases as we lower the temperature during the superconducting phase, 
similar to the result of $(2+1)$-dimensional holographic $\mathcal{O}_{1}$ order superconductor 
in Ref. \cite{Chakraborty:2019vld}.
It shows that the HSC has the potential to offer much richer physical information than the HEE,
although the physical mechanisms involved are still unclear.
It should be emphasized that we only focus on the ``CV'' conjecture of the subregion. 
It is interesting and necessary to use the ``CA'' conjecture and general ``CV'' conjecture to study
the holographic complexity in holographic superconducting systems.

\appendix
\section{the HEE and HSC for the normal phase} \label{Appendix}

In this appendix, we shall discuss the diverging term of
the HEE and HSC for a strip-shaped subregion $\mathcal{A}\subseteq \left(-L_x/2\leq x\leq L_x/2,-L_y/2\leq y\leq L_y/2  \right)$.
For simplicity, let us set $L=1$ for the rest of the calculation.
In the normal phase, the solution is simply the RN-AdS solution
\begin{eqnarray}
	f(z)=\frac{1}{z^2}-z+\frac{(z^2-z)\kappa ^2\rho ^2}{2},\       \ \phi(z)=(1-z)\rho.    
\end{eqnarray}
In the probe limit $\kappa ^2\rightarrow 0$, i.e., in the absence of backreaction, 
since the integration \eqref{HEE2} cannot be solved,
we can take subregion $L_x\rightarrow 0$, i.e., the minimal surface near the boundary $z\rightarrow 0$. 
In this case, the metric function has a simple behavior $f(z)\rightarrow 1/z^2$, and then the integral \eqref{HEE2} becomes integrable.
Therefore, the total HEE for a small subsystem in a pure AdS background is obtained  
\begin{eqnarray}
	\mathcal{S}=\frac{2 L_y}{4G_{N}} \int^{z_*}_{\epsilon }\frac{z^2_*dz}{z^3\sqrt{f(z^4_*-z^4)} }
	&=&\frac{ L_y}{2G_{N}} \int^{z_*}_{\epsilon }\frac{z^2_*dz}{z^3\sqrt{\frac{1}{z^2}\left(z^4_*-z^4\right)} }\nonumber\\
	&=&\frac{ L_y}{2G_{N}}\left( \frac{1}{\epsilon }-\frac{\sqrt{\pi}\Gamma(\frac{3}{4})}{z_*\Gamma(\frac{1}{4})}  \right),
\end{eqnarray}
where $1/\varepsilon $ is the divergent term caused by the pure AdS background, and $\Gamma$ is the gamma function.
Therefore, for all 4-dimensional asymptotic AdS black holes, the total HEE can be expressed in the general form 
\begin{eqnarray}
	\mathcal{S}=\frac{L}{2G_{N}}\left(\mathcal{S}_{u}+\frac{1}{\varepsilon }\right),
\end{eqnarray}
where $S_{u}$ is the universal term of total HEE and is physically important for asymptotic AdS black holes.
Considering the same approximation, we can also solve the integral of the minimal surface in the Eq. \eqref{minsur}  
\begin{eqnarray}
	x(z)&=&\int ^{z_*}_{z}\frac{zdz}{\sqrt{f(z^4_*-z^4)}}=\int ^{z_*}_{z}\frac{zdz}{\sqrt{\frac{1}{z^2}(z^4_*-z^4)}}\\
	&=&z_{*}\left\{EllipticE(-1)-EllipticK(-1)+EllipticF\left[ArcSin\left(\frac{z}{z_*}  \right) ,-1\right]\right\}\nonumber\\
	&&-z_{*}\left\{EllipticE\left[ArcSin\left(\frac{z}{z_*}  \right),-1 \right] \right\},\nonumber
\end{eqnarray}
where $EllipticE$, $EllipticK$, and $EllipticF$ are the second and
first complete elliptic integrals and first elliptic integrals, respectively.
Using Eq. \eqref{HSC2}, the total HSC for a small subsystem in a pure AdS background is obtained  
\begin{eqnarray}
	\mathcal{C}&=&\frac{2L_{y}}{8\pi L G_{N}}\int^{z_*}_{\epsilon }\frac{x(z)dz}{z^4\sqrt{f}}
	=\frac{L_{y}}{4\pi L G_{N}}\int^{z_*}_{\epsilon }\frac{x(z)dz}{z^4\sqrt{\frac{1}{z^2}}}   \nonumber  \\
	&\approx &\frac{L_{y}}{4\pi L G_{N}}\left\{\frac{z_*}{2\epsilon^2 }\left[EllipticE(-1)-EllipticK(-1)\right]-\frac{EllipticK(-1)}{2z_*}+\mathcal{O}(\epsilon )   \right\},
\end{eqnarray}
where $\frac{\mathcal{F}(z_{*})}{\varepsilon^2}$ is the divergent term of the total HSC caused by the pure AdS background.
For different cases, the numerators $\mathcal{F}(z_{*})$ of the divergent terms have different forms, but we can still get them numerically.
Therefore, for all 4-dimensional asymptotic AdS black holes, the total HSC can be expressed in the general form 
\begin{eqnarray}
	\mathcal{C}=\frac{L_{y}}{4\pi L G_{N}}\left[C_{u}+\frac{\mathcal{F}(z_{*})}{\varepsilon^2}   \right],
\end{eqnarray}
where $C_{u}$ is the universal term of the total HSC and is physically important for asymptotic AdS black holes.

\acknowledgments
We thank Prof. Qiu Taotao of Huazhong University of Science and Technology 
and Dr. Shi Jiaming of Hangzhou Institute for Advanced Study for their helpful discussions and suggestions.
This work was supported by the National Natural
Science Foundation of China under Grants No. 11653002 and No. 11875141,
and Postgraduate Scientific Research Innovation Project of Hunan Province (Grant No. CX20210472).

\bibliography{references}
\bibliographystyle{apsrev4-1}

\end{document}